\documentclass[pdflatex,sn-mathphys-num]{sn-jnl}

\usepackage{graphicx}%
\usepackage{multirow}%
\usepackage{amsmath,amssymb,amsfonts}%
\usepackage{amsthm}%
\usepackage{mathrsfs}%
\usepackage[title]{appendix}%
\usepackage{xcolor}%
\usepackage{textcomp}%
\usepackage{manyfoot}%
\usepackage{booktabs}%
\usepackage{algorithm}%
\usepackage{algorithmicx}%
\usepackage{algpseudocode}%
\usepackage{listings}%

\usepackage{color}

\theoremstyle{thmstyleone}%
\theoremstyle{thmstyletwo}%

\theoremstyle{thmstylethree}%

\begin{document}

\title[Quasiperiodic trajectories drawn by the Bloch vector of the thermal multiphoton Jaynes-Cummings model]{
Quasiperiodic trajectories drawn by the Bloch vector of the thermal multiphoton Jaynes-Cummings model
}

\author{\fnm{Hiroo} \sur{Azuma}}\email{zuma@nii.ac.jp}

\affil{\orgdiv{Global Research Center for Quantum Information Science}, \orgname{National Institute of Informatics}, \orgaddress{\street{2-1-2 Hitotsubashi, Chiyoda-ku}, \city{Tokyo}, \postcode{101-8430}, \country{Japan}}}

\abstract{
We study the time evolution of the Bloch vector of the thermal multiphoton Jaynes-Cummings model (JCM). If the multiphoton JCM incorporates thermal fluctuations, its corresponding Bloch vector evolves unpredictably, traces a disordered trajectory, and exhibits quasiperiodicity. However, if we plot the trajectory as a discrete-time sequence with a constant time interval, it reveals unexpected regularities. First, we show that this plot is invariant under a scale transformation of a finite but non-zero time interval. Second, we numerically evaluate the times at which the absolute value of the $z$-component of the Bloch vector is nearly equal to zero. At those times, the density matrix of the two-level system approximates a classical ensemble of the ground and excited states. We demonstrate that some time values can be derived from the denominators of the fractions of certain approximations for irrational numbers. The reason underlying these findings is that the components of the Bloch vector for the thermal multiphoton JCM are described with a finite number of trigonometric functions whose dimensionless angular frequencies are irrational numbers in the low-temperature limit.
}


\maketitle

\section{\label{section-introduction}Introduction}
The recent past has witnessed rapid progress in the technologies of quantum computation. Here, the goal of achieving practical quantum computations requires one to implement quantum bits (qubits) and quantum gates that generate entanglement between two qubits. Possible candidates for such qubits are superconducting Josephson-junction devices \cite{Arute2019, Osman2021}, linear trapped ions \cite{Kanshal2020,Murali2020}, quantum dots \cite{Veldhorst2014,Pino2024}, photons in linear optical circuits \cite{Liu2023,Yamazaki2024}, and so on. In association with the development of manufacturing technologies for qubits, the theoretical and experimental aspects of quantum optics have drawn the attention of many researchers in the field of quantum information science because quantum optics has a wide range of applications in quantum computation, teleportation, and cryptography.

In particular, the Jaynes-Cummings model (JCM) is a typical quantum optics system that has been studied since the 1980s \cite{Rempe1987,Lee2017}. It describes a simple situation where the raising and lowering operators of an atom are coupled to single annihilation and creation operators of photons. Because it is a fully quantum and solvable system, it has often been used as a convenient test bed for studying the quantum properties of atoms and cavity fields of photons.

The JCM is used for many purposes in quantum information processing. For example, the JCM can be incorporated in the Hamiltonian of a strongly coupled atom-cavity system for controlled generation of single photons \cite{Kuhn1999}. Ion-trap quantum computing models the atom-cavity interaction with the JCM, as well \cite{Cirac1995,Steane1997}. In regard to transmon qubits which consist of two superconducting islands coupled via two Josephson junctions, we can regard the Hamiltonian of the transmon and modes of the LC resonator to be a JCM \cite{Larson2021}.

A multiphoton JCM is a natural extension of the JCM \cite{Sukumar1981,Singh1982,Sukumar1984,Shumovsky1985}.
It has an interaction Hamiltonian which allows raising and lowering operators of a single atom (that is, $\sigma_{\pm}$) to be coupled to the annihilation and creation operators of $l$ photons [that is, $a^{l}$ and $(a^{\dagger})^{l}$], respectively, for $l=2,3,...$. Particular implementations of two-photon JCMs have been discussed in the form of a single atom inside an optical cavity \cite{Zou2020,Tang2023} and superconducting circuits \cite{Felicetti2018}. Moreover, a two-photon JCM has been theoretically derived from the two-photon quantum Rabi model by applying the rotating-wave approximation, and a scheme for realizing one with trapped ions has been proposed \cite{Puebla2017}. In addition, two-mode multiphoton JCMs have recently been investigated and the quantum properties of noisy cases and analytical solutions of ideal noiseless cases have been derived \cite{Laha2024a,Laha2024b}.

The multiphoton JCMs exhibit different physical properties from those of single-photon JCMs in various situations. Two-photon JCMs have been shown to be more insensitive to thermal noise than single-photon JCMs \cite{Azuma2024}. In \cite{Azuma2024}, the initial state of the system was treated as a product of the ground state of the atom and the coherent state of the cavity field: the period spanning the collapse and revival of the Rabi oscillations and the relative entropy of coherence were calculated up to a second-order perturbation of the low-temperature expansion. The phase diagram of the two-photon Jaynes-Cummings-Hubbard model (JCHM) has been shown to be different from that of the single-photon JCHM \cite{Azuma2025}. The JCHM is a lattice of coupled high-Q microcavities, each of which is described by a JCM. The single- and multiphoton JCHMs have revealed quantum phase transitions between Mott-insulator and superfluid phases. Reference~\cite{Zou2020} studied the multiphoton blockade in a two-photon JCM whose behavior was different from that in a single-photon JCM. The above facts motivated us to explore the multiphoton JCMs.

The Bloch vector is a three-dimensional real vector whose length is equal to or less than unity. Because there is a one-to-one correspondence between the Bloch vector and a state of a two-level system, the Bloch vector is a very convenient way to visualize a quantum state. If the length of the Bloch vector is equal to unity, the system is in a pure state. Contrastingly, if the length is less than unity, the system is in a mixed state. We give an explicit definition of the Bloch vector in Sec.~\ref{section-Bloch-vector-thermal-multiphoton-JCM}.

In this paper, we study the time evolution of the Bloch vector of an $l$-photon JCM, $l=2,3,4$, where the photons are in contact with a thermal reservoir. We show that if the initial state consists of an atom in a superposition of the ground and excited states and a cavity field in a mixed state obeying a Bose-Einstein distribution, the time evolution of the Bloch vector under the JCM interaction is disordered. However, if we draw the trajectory of the Bloch vector not for a continuous time variable but rather for a sequence of discrete time points, it reveals a characteristic regularity. As explained in Sec.~\ref{section-Bloch-vector-thermal-multiphoton-JCM}, the sequence of discrete times bestows quasiperiodicity on the time evolution of the Bloch vector of the multiphoton JCM \cite{Ott1993}. This notion of quasiperiodicity allows us to regard the trajectory of the Bloch vector to be an intermediate motion between periodic and chaotic. That is, quasiperiodicity underpins our choice of a discrete-time variable over the continuous one.

We made two findings in this study. In particular, we plotted the trajectory of the time evolution of the Bloch vector $\mbox{\boldmath $S$}(t)$ as a discrete sequence of points on the $xz$-plane $\{(S_{x}(t_{n}),S_{z}(t_{n})): t_{n}=n\Delta t,n=0,1,2,...,N\}$ for large $N$ and found that the obtained graph is invariant under a scale transformation $\Delta t\to s\Delta t$, where $s$ is an arbitrary real but not transcendental number and $\pi<s\Delta t (\mbox{mod $2\pi$})$ with $\pi<\Delta t (\mbox{mod $2\pi$})$.
This implies that lower and upper bounds of $s$ exist and they are given by
\begin{equation}
\frac{\pi}{\Delta t(\mbox{mod $2\pi$})}
<
s
<
\frac{2\pi}{\Delta t(\mbox{mod $2\pi$})}
\end{equation}
Next, we numerically evaluated the times at which the absolute value of the $z$-component of the Bloch vector is nearly equal to zero, i.e., $|S_{z}(t)|<\epsilon$ for a small positive number $\epsilon(\ll 1)$. We found that we could compute some of those time values from the denominators of fractions of the Diophantine approximations of irrational numbers. Although our previous study showed that these two findings apply to the original JCM in \cite{Azuma2014}, the current study shows that they also hold for the multiphoton JCM.

The reason underlying the above two findings is that the components of the Bloch vector are described with a finite number of trigonometric functions whose dimensionless angular frequencies are irrational numbers in the low-temperature limit.
We derive closed mathematical forms of the components of the Bloch vector which include terms of $\cos(\sqrt{n}t)$ and $\sin(\sqrt{n}t)$, where $n$ denotes the summation index in cases having finite but not zero temperature. (Strictly speaking, $n$ represents the number of thermal photons in the Bose-Einstein distribution at finite temperature.) Thus, the above two findings essentially depend on thermal effects.

This paper is organized as follows. In Sec.~\ref{section-Bloch-vector-thermal-multiphoton-JCM}, we derive closed mathematical expressions for the Bloch vector of the multiphoton JCMs. In Sec.~\ref{section-discrete-plot-Bloch-vector}, we draw discrete plots of the time evolution of the Bloch vector and show that they are invariant under scale transformations of the time variable. In Sec.~\ref{section-Sz-zero-time}, we numerically calculate times at which the absolute value of the $z$-component of the Bloch vector is nearly equal to zero and plot $(\beta,S_{x}(t))$ for time $t$ such that $|S_{z}(t)|<\epsilon$. Then, we show that the values of $t$ can be derived from Diophantine approximations for an example of the two-photon JCM. Section~\ref{section-conclusion-discussion} is a conclusion that summarizes the findings and mentions future work. Appendix~\ref{section-appendix-A} presents how to implement the two-photon JCM experimentally using Josephson junctions. Appendix~\ref{section-appendix-B} gives details of the computation of the time $t$ such that $|S_{z}(t)|<\epsilon$ for $l$-photon JCMs with $l=1$, $3$, and $4$ by using Diophantine approximations.

\section{\label{section-Bloch-vector-thermal-multiphoton-JCM}The Bloch vector of the multiphoton JCM}
The Hamiltonian of the original and the multiphoton JCM is given by
\begin{equation}
H
=
\frac{\omega_{0}}{2}\sigma_{z}
+
\omega a^{\dagger}a
+
g[\sigma_{+}a^{l}+\sigma_{-}(a^{\dagger})^{l}]
\quad
\mbox{for $l=1,2,3,...$},
\label{multiphoton-JCM-Hamiltonian-0}
\end{equation}
where
\begin{equation}
\sigma_{x}
=
\left(
\begin{array}{cc}
0 & 1 \\
1 & 0 \\
\end{array}
\right),
\quad
\sigma_{y}
=
\left(
\begin{array}{cc}
0 & -i \\
i & 0 \\
\end{array}
\right),
\quad
\sigma_{z}
=
\left(
\begin{array}{cc}
1 & 0 \\
0 & -1 \\
\end{array}
\right),
\end{equation}
$\sigma_{\pm}=(1/2)(\sigma_{x}\pm i\sigma_{y})$, the orthonormal basis of the atom is represented as two-component vectors,
\begin{equation}
|0\rangle_{\text{A}}
=
\left(
\begin{array}{c}
1 \\
0 \\
\end{array}
\right),
\quad
|1\rangle_{\text{A}}
=
\left(
\begin{array}{c}
0 \\
1 \\
\end{array}
\right),
\end{equation}
and $a$ and $a^{\dagger}$ are annihilation and creation operators of photons, respectively \cite{Walls2008}. We set $\hbar=1$ and denote the angular frequencies of the atom and photons as $\omega_{0}$ and $\omega$, respectively. The dimension of the coupling constant $g$ is equal to that of the frequency. Moreover, for simplicity, we assume that $g$ is a real constant. Here, we must set $l$ to be a certain integer in order to perform concrete calculations. How to implement the two-photon JCM experimentally is provided in Appendix~\ref{section-appendix-A}.

Now, we split up $H$ as follows:
\begin{eqnarray}
H
&=&
C_{1}+C_{2}, \nonumber \\
C_{1}
&=&
\omega[(l/2)\sigma_{z}+a^{\dagger}a], \nonumber \\
C_{2}
&=&
-(\Delta\omega/2)\sigma_{z}
+
g[\sigma_{+}a^{l}+\sigma_{-}(a^{\dagger})^{l}],
\label{C1-C2-definitions-0}
\end{eqnarray}

\begin{equation}
\Delta\omega=-\omega_{o}+l\omega.
\end{equation}
We obtain $[C_{1},C_{2}]=0$. This commutation relation leads to the interaction picture, as follows. First, we describe the wave function in the Schr{\"{o}}dinger picture as $|\Psi_{\text{S}}(t)\rangle$. Second, we introduce a new wave function in the interaction picture,
\begin{equation}
|\Psi_{\text{I}}(t)\rangle
=
e^{iC_{1}t}|\Psi_{\text{S}}(t)\rangle.
\label{interaction-picture-wave-function-0}
\end{equation}
Thus, we can write the state of the system in the interaction picture as
\begin{equation}
|\Psi_{\text{I}}(t)\rangle
=
U(t)
|\Psi_{\text{I}}(0)\rangle,
\end{equation}
\begin{eqnarray}
U(t)
&=&
\exp(-iC_{2}t) \nonumber \\
&=&
\left(
\begin{array}{cc}
u_{00} & u_{01} \\
u_{10} & u_{11} \\
\end{array}
\right).
\label{definition-unitary-C2-0}
\end{eqnarray}

From here on, we will assume $\Delta\omega=0$ for the sake of simplicity. This implies that we set $\omega_{0}$ and $\omega$ so that they satisfy $\omega_{0}=l\omega$. This simplification will play an important role in deriving time values satisfying $|S_{z}(t)|<\epsilon$ for $0<\epsilon\ll 1$ with the Diophantine approximation in Sec.~\ref{section-Sz-zero-time}, where $S_{z}(t)$ is the $z$-component of the Bloch vector.

The following are explicit forms of the elements of the matrix $U(t)$:
\begin{eqnarray}
u_{00}
&=&
\cos(\sqrt{D}t), \nonumber \\
u_{01}
&=&
-ig
\frac{\sin(\sqrt{D}t)}{\sqrt{D}}
a^{l}, \nonumber \\
u_{10}
&=&
-ig
\frac{\sin(\sqrt{D'}t)}{\sqrt{D'}}
(a^{\dagger})^{l}, \nonumber \\
u_{11}
&=&
\cos(\sqrt{D'}t),
\label{matrix-elements-unitary-0}
\end{eqnarray}

\begin{eqnarray}
D
&=&
g^{2}a^{l}(a^{\dagger})^{l}, \nonumber \\
D'
&=&
g^{2}(a^{\dagger})^{l}a^{l}.
\label{D-D-dash-operators-0}
\end{eqnarray}
We set the initial state of the atom and photons as $\rho_{\text{AP}}(0)=\rho_{\text{A}}(0)\otimes\rho_{\text{P}}$. Thus, $\rho_{\text{AP}}(0)$ is a state in which the atom and photons are not entangled. We define the matrix elements of $\rho_{\text{A}}(0)$ as follows:
\begin{equation}
\rho_{\text{A}}(0)
=
\sum_{i,j\in\{0,1\}}
\rho_{\text{A},ij}(0)|i\rangle_{\text{A}}{}_{\text{A}}\langle j|.
\end{equation}
In addition, we assume that the photons are initially in contact with a thermal reservoir and they are in thermal equilibrium at temperature $T$. Thus, the density operator of the photons at the initial time $\rho_{\text{P}}$ is given by the Bose-Einstein distribution,
\begin{equation}
\rho_{\text{P}}
=
(1-e^{-\beta\omega})\exp(-\beta\omega a^{\dagger}a),
\label{Bose-Einstein-distribution-0}
\end{equation}
where $\beta=1/k_{\text{B}}T$ and $k_{\text{B}}$ denotes the Boltzmann constant. Thus, we have introduced the temperature into the multiphoton JCM through Eq.~(\ref{Bose-Einstein-distribution-0}).

The time evolution of the state of the atom is given by
\begin{eqnarray}
\rho_{\text{A}}(t)
&=&
\mbox{Tr}_{\text{P}}[U(t)\rho_{\text{AP}}(0)U^{\dagger}(t)] \nonumber \\
&=&
\sum_{k,m\in\{0,1\}}
\rho_{\text{A},km}(t)|k\rangle_{\text{A}}{}_{\text{A}}\langle m|,
\end{eqnarray}
where
\begin{equation}
\rho_{\text{A},km}(t)
=
\sum_{i,j\in\{0,1\}}
\rho_{\text{A},ij}(0)A_{ij,km}(t)
\quad
\mbox{for $k,m\in\{0,1\}$},
\end{equation}

\begin{equation}
A_{ij,km}(t)
=
{}_{\text{A}}\langle k|
\mbox{Tr}_{\text{P}}
[U(t)(|i\rangle_{\text{A}}{}_{\text{A}}\langle j|\otimes\rho_{\text{P}})
U^{\dagger}(t)]|m\rangle_{\text{A}}.
\end{equation}
The matrix elements of $\rho_{\text{A}}(t)$ satisfy the relationships, $\rho_{\text{A},10}(t)=\rho_{\text{A},01}(t)^{*}$ and \\ $\rho_{\text{A},11}(t)=1-\rho_{\text{A},00}(t)$. Now, we only have to compute $\rho_{\text{A},00}(t)$ and $\rho_{\text{A},01}(t)$. The matrix elements $A_{ij,km}(t)$ have the following forms:
\begin{eqnarray}
A_{00,00}(t)
&=&
(1-e^{-\beta\omega})\sum_{n=0}^{\infty}e^{-n\beta\omega}
\cos^{2}(\sqrt{D_{n}}t), \nonumber \\
A_{11,00}(t)
&=&
g^{2}(1-e^{-\beta\omega})
\sum_{n=0}^{\infty}
\frac{\sin^{2}(\sqrt{D_{n}}t)}{D_{n}}
\prod_{k=1}^{l}(n+k)
e^{-(n+l)\beta\omega}, \nonumber\\
A_{01,01}(t)
&=&
(1-e^{-\beta\omega})
\sum_{n=0}^{\infty}
\cos(\sqrt{D_{n}}t)
\cos(\sqrt{D_{n}'}t)
e^{-n\beta\omega},
\label{A-elements-0}
\end{eqnarray}
where $D_{n}$ and $D_{n}'$ are eigenvalues of $D$ and $D'$ such as $D|n\rangle_{\text{P}}=D_{n}|n\rangle_{\text{P}}$, $D'|n\rangle_{\text{P}}=D_{n}'|n\rangle_{\text{P}}$, $\{|n\rangle_{\text{P}}:n=0, 1, 2, ...\}$:
\begin{equation}
D_{n}=g^{2}\prod_{k=1}^{l}(n+k),
\label{definition-D-n}
\end{equation}

\begin{equation}
D'_{n}
=
\left\{
\begin{array}{ll}
g^{2}\prod_{k=1}^{l}(n-k+1) & \mbox{for $n\geq l$}, \\
0 & \mbox{for $n\leq l-1$}, \\
\end{array}
\right.
\label{definition-D-dash-n}
\end{equation}
and $A_{01,00}(t)=A_{10,00}(t)=A_{00,01}(t)=A_{10,01}(t)=A_{11,01}(t)=0$. Accordingly, we have
\begin{eqnarray}
\rho_{\text{A},00}(t)
&=&
\rho_{\text{A},00}(0)A_{00,00}(t)
+
\rho_{\text{A},11}(0)A_{11,00}(t), \nonumber \\
\rho_{\text{A},01}(t)
&=&
\rho_{\text{A},01}(0)A_{01,01}(t).
\end{eqnarray}
By introducing the Bloch vector $\mbox{\boldmath $S$}(t)=(S_{x}(t),S_{y}(t),S_{z}(t))$,
\begin{equation}
\rho_{\text{A}}(t)
=
\frac{1}{2}
[
\mbox{\boldmath $I$}+\mbox{\boldmath $S$}(t)\cdot\mbox{\boldmath $\sigma$}],
\end{equation}
where $\mbox{\boldmath $\sigma$}=(\sigma_{x},\sigma_{y},\sigma_{z})$, we obtain
\begin{equation}
\mbox{\boldmath $S$}(t)
=
\left(
\begin{array}{ccc}
L^{(1)}(t) & 0 & 0 \\
0 & L^{(1)}(t) & 0 \\
0 & 0 & L^{(2)}(t) \\
\end{array}
\right)
\mbox{\boldmath $S$}(0)
+
\left(
\begin{array}{c}
0 \\
0 \\
L^{(3)}(t) \\
\end{array}
\right),
\end{equation}

\begin{eqnarray}
L^{(1)}(t)
&=&
A_{01,01}(t), \nonumber \\
L^{(2)}(t)
&=&
A_{00,00}(t)-A_{11,00}(t), \nonumber \\
L^{(3)}(t)
&=&
A_{00,00}(t)+A_{11,00}(t)-1.
\label{L1234-definition}
\end{eqnarray}

Moreover, setting the initial state of the atom to $\rho_{\text{A}}(0)=|\psi\rangle_{\text{A}}{}_{\text{A}}\langle\psi|$ and $|\psi\rangle_{\text{A}}=(1/\sqrt{2})(|0\rangle_{\text{A}}+|1\rangle_{\text{A}})$, we can express the initial Bloch vector as $\mbox{\boldmath $S$}(0)=(1,0,0)^{\text{T}}$. Finally, we attain
\begin{equation}
\mbox{\boldmath $S$}(t)
=
\left(
\begin{array}{c}
L^{(1)}(t) \\
0 \\
L^{(3)}(t) \\
\end{array}
\right).
\label{Bloch-vector-xz-plane-0}
\end{equation}
Thus, in this circumstance, we only need to consider Bloch vectors on the $xz$-plane.

\begin{figure}
\begin{center}
\includegraphics[width=0.45\linewidth]{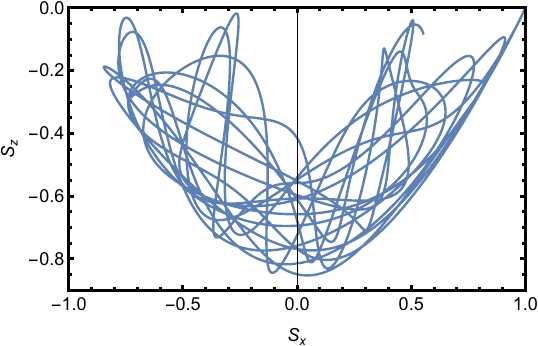}
\end{center}
\caption{
A trajectory of $\mbox{\boldmath $S$}(t)$ given by Eqs.~(\ref{L1234-definition}) and (\ref{Bloch-vector-xz-plane-0}) for the two-photon JCM ($l=2$) with $g=\omega=\beta=1$. The time variable is continuous in the range $0\leq t\leq 40$. The trajectory is disordered.
}
\label{figure01}
\end{figure}

Figure~\ref{figure01} shows a plot of the trajectory of $\mbox{\boldmath $S$}(t)$ for a two-photon JCM with a continuous-time variable $0\leq t\leq 40$ and $g=\omega=\beta=1$. Because of thermal effects, the behavior of the Bloch vector in Fig.~\ref{figure01} seems to be completely disordered.

The time evolution of the Bloch vector $\mbox{\boldmath $S$}(t)$ given by Eqs.~(\ref{L1234-definition}) and (\ref{Bloch-vector-xz-plane-0}) shows quasiperiodicity, an intermediate motion between periodic and chaotic ones. The following considerations can help us to understand this phenomenon. For the single-photon JCM ($l=1$) in the low-temperature limit ($\beta\gg 1$), an approximate form of $L^{(1)}(t)$ is given by
\begin{equation}
L^{(1)}(t)
\simeq
(1-e^{-\beta\omega})\cos t
+
e^{-\beta\omega}\cos t\cos(\sqrt{2}t),
\end{equation}
by neglecting terms of order $O[(e^{-\beta\omega})^{2}]$ and letting $g=1$ for simplicity. On the one hand, the first term of the above equation includes $\cos t$. On the other hand, the second term includes $\cos t \cos(\sqrt{2}t)$. Thus, $L^{(1)}(t)$ is not periodic in the time variable $t$. However, the Diophantine approximation tells us that there are an infinite number of pairs $(p,q)$ where $p$ and $q$ are coprime and satisfy the relationship \cite{Coppel2009,Niven1960},
\begin{equation}
\left|
\sqrt{2}-\frac{p}{q}
\right|
<
\frac{1}{q^{2}}.
\label{root-2-Diophantine-approximation-0}
\end{equation}
Thus, we find that
\begin{equation}
\cos(\sqrt{2}t)\simeq\cos
[(p/q)t],
\end{equation}
and $\cos(\sqrt{2}t)\simeq\cos(2p\pi)=1$ holds for $t=2q\pi$. Hence, we find that $L^{(1)}(0)\simeq L^{(1)}(2q\pi)$ and reach the following conclusion. Although $L^{(1)}(t)$ is not periodic in the time variable $t$, if we choose $q$ satisfying Eq.~(\ref{root-2-Diophantine-approximation-0}) as large as possible, we can let $L^{(1)}(2q\pi)$ be close to $L^{(1)}(0)$ to arbitrary precision. This property is called quasiperiodicity. We have developed the above considerations for $l=1$. Moreover, a similar discussion applies to the cases of $l=2$, $3$, and $4$. Hence, the time evolution of the Bloch vector $\mbox{\boldmath $S$}(t)$ for the multiphoton JCM has quasiperiodicity. However, the trajectory of $\mbox{\boldmath $S$}(t)$ for the continuous time variable is disordered when the duration is not long enough, as shown in Fig.~\ref{figure01}.

\section{\label{section-discrete-plot-Bloch-vector}Discrete plots of the time evolution of the Bloch vector}
\begin{figure}
\begin{center}
\includegraphics[width=\linewidth]{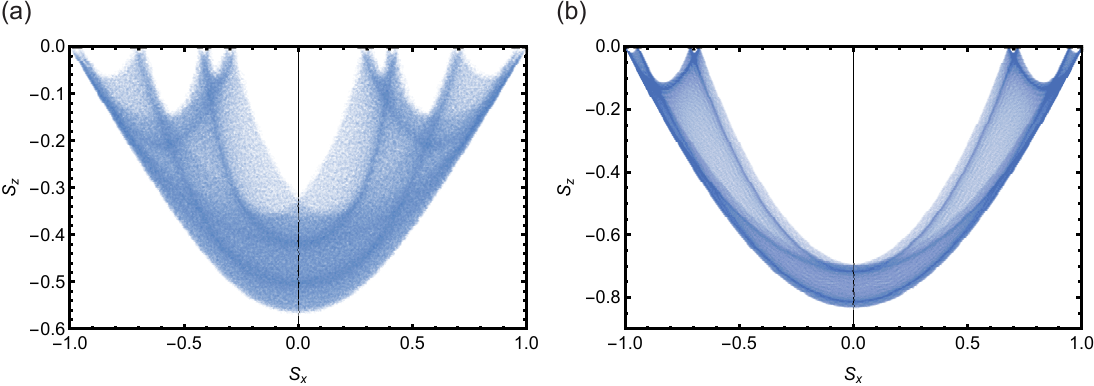}
\end{center}
\caption{Graphs consisting of points of $\mbox{\boldmath $S$}(t)$ given by Eqs.~(\ref{L1234-definition}) and (\ref{Bloch-vector-xz-plane-0}) for the original JCM ($l=1$). The points are plotted at constant time intervals $\Delta t=4$ for $\omega=1$ and $g=1$. The total number of points is $N=400{\,}001$. The plots in Figs.~\ref{figure03}, \ref{figure04}, and \ref{figure05} use the same parameters. (a) $\beta=0.9$. (b) $\beta=1.8$.}
\label{figure02}
\end{figure}

\begin{figure}
\begin{center}
\includegraphics[width=\linewidth]{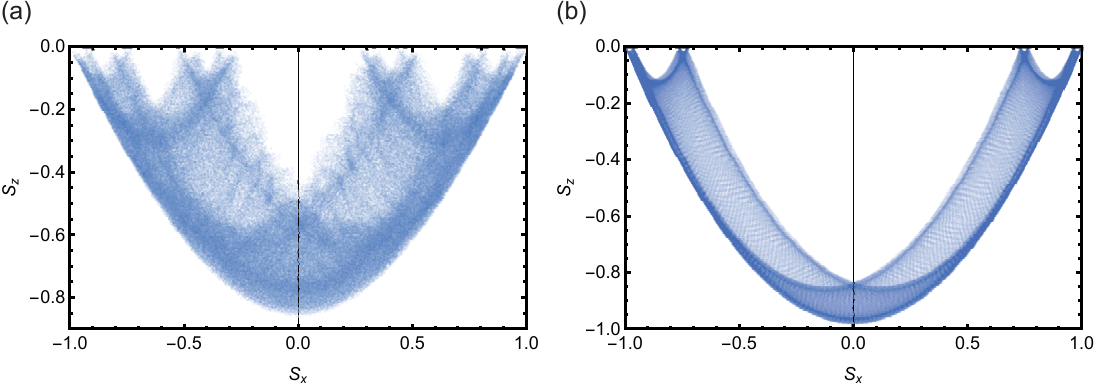}
\end{center}
\caption{Graphs consisting of points of $\mbox{\boldmath $S$}(t)$ for the two-photon JCM ($l=2$). (a) $\beta=1.0$. (b) $\beta=2.0$.}
\label{figure03}
\end{figure}

\begin{figure}
\begin{center}
\includegraphics[width=\linewidth]{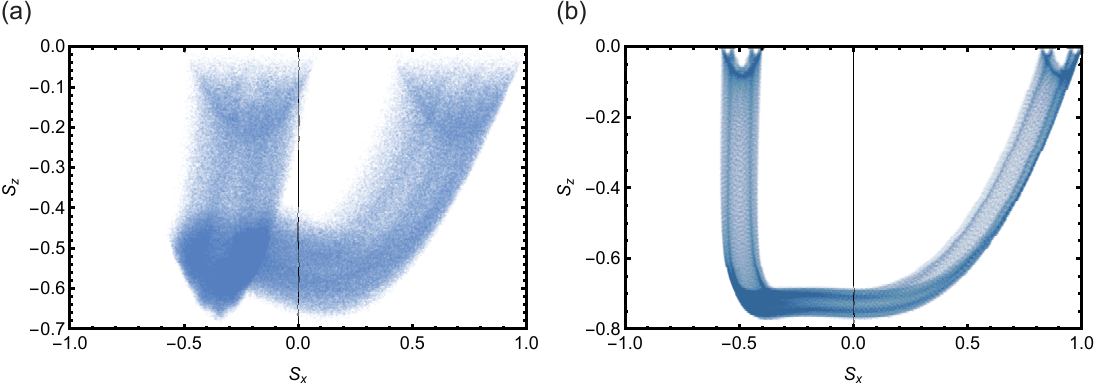}
\end{center}
\caption{Graphs consisting of points of $\mbox{\boldmath $S$}(t)$ for the three-photon JCM ($l=3$). (a) $\beta=0.6$. (b) $\beta=1.2$.}
\label{figure04}
\end{figure}

\begin{figure}
\begin{center}
\includegraphics[width=\linewidth]{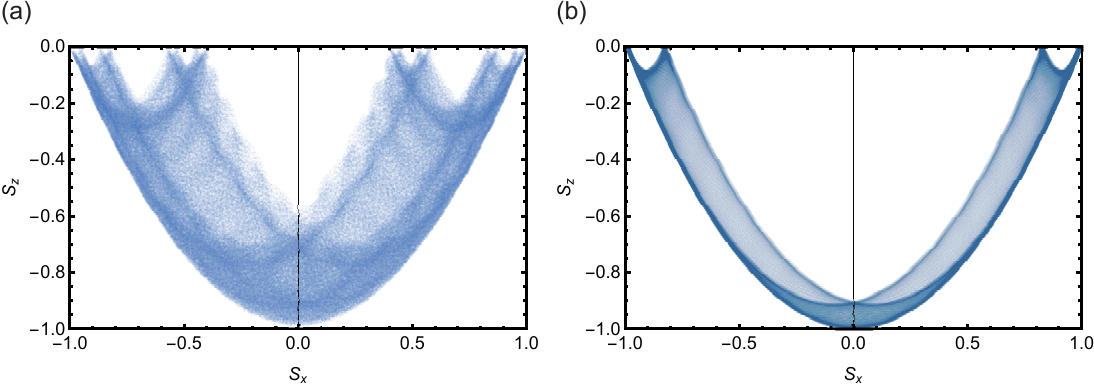}
\end{center}
\caption{Graphs consisting of points of $\mbox{\boldmath $S$}(t)$ for the four-photon JCM ($l=4$). (a) $\beta=1.2$. (b) $\beta=2.4$.}
\label{figure05}
\end{figure}

Here, we examine the graphs $\mbox{\boldmath $S$}(t)$ produced by Eqs.~(\ref{L1234-definition}) and (\ref{Bloch-vector-xz-plane-0}). Figures~\ref{figure02}, \ref{figure03}, \ref{figure04}, and \ref{figure05} plot $\mbox{\boldmath $S$}(t)$ at constant time intervals $t_{n}=n\Delta t$ for $n=0,1,2,...,N$ with large $N$ on the $xz$-plane.

The graphs in Figs.~\ref{figure02}, \ref{figure03}, \ref{figure04}, and \ref{figure05} are invariant under a scale transformation $\Delta t\to s\Delta t$, where $s$ is an arbitrary real but not transcendental number and $\pi<s\Delta t(\mbox{mod $2\pi$})$ with $\pi<\Delta t(\mbox{mod $2\pi$})$. Here, we explain why the above statement holds. First, let us consider a set of $N$ numbers, ${\cal S}_{N}=\{x_{j}:0\leq x_{j}<2\pi,j\in\{0,1,...,N\}\}$. The necessary and sufficient condition that the sequence $(x_{0},x_{1},...,x_{N})$ is uniformly distributed in the range of $[0,2\pi)$ under the limit of $N\to +\infty$ is as follows \cite{Coppel2009,Weyl1916,Davenport1963,Kuipers1974}:
\begin{equation}
\lim_{N\to +\infty}
\frac{1}{N+1}
\sum_{j=0}^{N}
\exp(imx_{j})=0
\quad
\forall m\in\{\pm 1,\pm 2,...\}.
\end{equation}
Here, we let ${\cal S}_{N}=\{x_{n}=n\Delta t (\mbox{mod $2\pi$}): n\in\{0,1,...,N\}\}$. Accordingly, we have
\begin{equation}
\lim_{N\to +\infty}
\frac{1}{N+1}
\sum_{n=0}^{\infty}
\exp(imx_{n})
=
\lim_{n\to +\infty}
\frac{1}{N+1}
\frac{1-e^{im(N+1)\Delta t}}{1-e^{im\Delta t}}.
\label{condition-uniformly-distributed-sequence}
\end{equation}
Thus, if $m\Delta t (\mbox{mod $2\pi$})\neq 0$ $\forall m\in\{\pm 1,\pm 2,...\}$, the right-hand side of Eq.~(\ref{condition-uniformly-distributed-sequence}) is equal to zero. Hence, if $\Delta t$ is a real, positive, but not transcendental number, $m\Delta t (\mbox{mod $2\pi$})\neq 0$ $\forall m$ holds and ${\cal S}_{N}$ becomes a uniformly distributed set. Second, let us assume $\pi<\Delta t$($\mbox{mod $2\pi$})$. Then, whenever the sequence number progresses from $n\Delta t$ to $(n+2)\Delta t$, congruence modulo $2\pi$ certainly occurs. Thus, $(x_{0},x_{1},x_{2},...,x_{N})$ becomes a pseudorandom sequence produced by a linear congruential generator.

From the above, we can regard $(0,\Delta t,2\Delta t,...,N\Delta t)$ to be a pseudorandom sequence uniformly distributed in the range $[0,2\pi)$. The components of $\mbox{\boldmath $S$}(t)$, i.e., $L^{(1)}(t)$ and $L^{(3)}(t)$, depend on $t$ via terms $\cos(\sqrt{D_{n}}t)$, $\cos(\sqrt{D'_{n}}t)$, and $\sin(\sqrt{D_{n}}t)$, where $D_{n}$ and $D'_{n}$ are given by Eqs.~(\ref{definition-D-n}) and (\ref{definition-D-dash-n}) with $\omega=1$ and $g=1$, i.e.,
\begin{equation}
D_{n}=\prod_{k=1}^{l}(n+k),
\label{definition-D-n-2}
\end{equation}

\begin{equation}
D'_{n}
=
\left\{
\begin{array}{ll}
\prod_{k=1}^{l}(n-k+1) & \mbox{for $n\geq l$}, \\
0 & \mbox{for $n\leq l-1$}. \\
\end{array}
\right.
\label{definition-D-dash-n-2}
\end{equation}
Further, $\sqrt{D_{n}}$ and $\sqrt{D'_{n}}$ are real, positive, but not transcendental. Thus, we cannot distinguish a graph that consists of $(L^{(1)}(t),L^{(3)}(t))$ for $t\in{\cal S}_{N}=\{n\Delta t:n=0,1,...,N\}$ from one that consists of $(L^{(1)}(t'),L^{(3)}(t'))$ for $t'\in{\cal S}'_{N}=\{n\Delta t':n=0,1,...,N\}$ with $\Delta t'=s\Delta t$ in the limit of $N\to +\infty$. Hence, we can conclude that the graph of $\{(L^{(1)}(t),L^{(3)}(t)):t=n\Delta t,n\in\{0,1,...,N\}\}$ is invariant under the scale transformation $\Delta t\to s\Delta t$ for $\pi<s\Delta t$($\mbox{mod $2\pi$}$) with $\pi<\Delta t$($\mbox{mod $2\pi$}$).

Looking at Figs.~\ref{figure02}, \ref{figure03}, and \ref{figure05} for $l=1$, $2$, and $4$, we see that the graphs are symmetrical with respect to the $S_{z}$-axis but not symmetrical with respect to the $S_{x}$-axis. The reasons are as follows. First, in the case of the Hamiltonian given by Eq.~(\ref{multiphoton-JCM-Hamiltonian-0}), we must determine two orthonormal vectors $\{|0\rangle_{\text{A}},|1\rangle_{\text{A}}\}$ as the ground and excited states of the atom. Accordingly, we select a specific direction for the $S_{z}$-axis of the Bloch vector. This choice breaks the symmetry of the inversion $S_{z}\to -S_{z}$. Second, we consider simultaneous inversions of the $S_{x}$- and $S_{y}$-axes. These inversions cause transformations $\sigma_{x}\to -\sigma_{x}$ and $\sigma_{y}\to -\sigma_{y}$ and they correspond to transformations $\sigma_{+}\to -\sigma_{+}$ and $\sigma_{-}\to -\sigma_{-}$. However, we can absorb the effects caused by these transformations by replacing the coupling constant $g$ with $-g$. Now, let us concentrate on Eqs.~(\ref{A-elements-0}), (\ref{definition-D-n}), and (\ref{definition-D-dash-n}). In these equations, the Bloch vector depends on $g^{2}$, but not $g$. Hence, the replacement of $g$ with $-g$ does not change the Bloch vector. Thus, the Bloch vector is invariant under simultaneous inversions of the $S_{x}$- and $S_{y}$-axes. From Eq.~(\ref{Bloch-vector-xz-plane-0}), the $S_{y}$-component of the Bloch vector is equal to zero $\forall t$. Thus, the Bloch vector is invariant for the inversion of $S_{x}$-axis, so that it is symmetrical with respect to the $S_{z}$-axis.

However, looking at Fig.~\ref{figure04} for the case $l=3$, we see that the graph is not symmetrical with respect to the $S_{z}$-axis. The reason for this is explained at the end of Sec.~\ref{section-Sz-zero-time}.

\section{\label{section-Sz-zero-time}Derivation of the time $t$ for $|S_{z}(t)|\ll 1$ with the Diophantine approximation}
\begin{figure}
\begin{center}
\includegraphics[width=\linewidth]{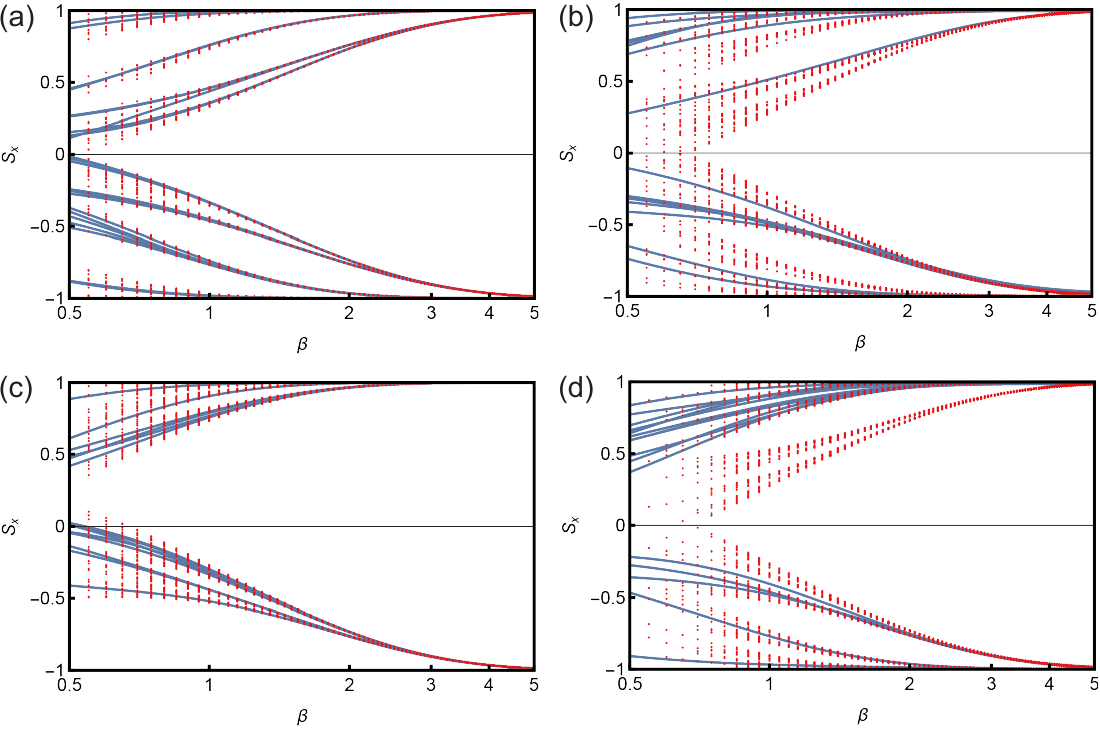}
\end{center}
\caption{The red points represent $(\beta,S_{x}(t_{n}))$ for $t_{n}=n\Delta t$ where $|S_{z}(t_{n})|<\epsilon$, where $\epsilon$ is a small positive number ($0<\epsilon\ll 1$). In calculating $S_{x}(t_{n})$ and $S_{z}(t_{n})$, we set $\Delta t=4$, $g=1$, and $\omega=1$. The horizontal axis is a logarithmic scale. For $2\leq\beta\leq 5.0$, we set $\epsilon=\epsilon_{0}$. By contrast, for $0.5\leq\beta<2.0$, we put $\epsilon=\epsilon_{0}\exp[c_{0}(2.0-\beta)]$. Selecting $t_{n}$ that satisfies $|S_{z}(t_{n})|<\epsilon$, we try $t_{n}=n\Delta t$ for $n=0,1,...,N$, where $N$ is a large number. We set $N$ as follows. For $0.5\leq\beta<1.0$, $1.0\leq\beta<2.0$, $2.0\leq\beta<3.0$, and $3.0\leq\beta\leq 5.0$, we set $N=N_{0}$, $N=N_{0}/2$, $N=N_{0}/5$, $N=N_{0}/10,$ respectively, where $N_{0}=1{\,}000{\,}000$. The blue curves represent $S_{x}(t)$ with $t=q\pi$ and $q\in\tilde{\cal M}$ given by Eq.~(\ref{tilde-M-l-2}) and Appendix~\ref{section-appendix-B}. Note that not all of the red points lie on the blue curves. (a) $l=1$ (the original JCM), $\epsilon_{0}=0.0035$, $c_{0}=0.7$. (b) $l=2$ (the two-photon JCM), $\epsilon_{0}=0.009$, $c_{0}=0.6$. (c) $l=3$ (the three-photon JCM), $\epsilon_{0}=0.0024$, $c_{0}=1.5$. (d) $l=4$ (the four-photon JCM), $\epsilon_{0}=0.009$, $c_{0}=0.7$.}
\label{figure06}
\end{figure}

The red points in Fig.~\ref{figure06} plot $S_{x}(t_{n})$ as a function of $\beta$ for discrete-time values $t_{n}=n\Delta t$, where $|S_{z}(t_{n})|<\epsilon$ holds for a small positive number $0<\epsilon\ll 1$, $l=1,2,3,4$.

Now, let us consider how to derive the value of the time variable $t$ that satisfies $|S_{z}(t)|<\epsilon$ analytically. Assuming $g=1$ and $\omega=1$, we can rewrite $L^{(3)}(t)$ as
\begin{equation}
L^{(3)}(t)
=
-\frac{1}{2}(1-e^{-\beta l})
[1-(1-e^{-\beta})\sum_{n=0}^{\infty}e^{-n\beta}
\cos(2\sqrt{D_{n}}t)].
\label{L-4-t}
\end{equation}
Hence, $L^{(3)}(t)=0$ holds $\forall \beta$ if $\cos(2\sqrt{D_{n}}t)=1$ $\forall n$.

First, let us consider the case where $n=0$. Because $D_{0}=l!$ from Eq.~(\ref{definition-D-n-2}), what we must compute is a real value of the time variable $t$ that satisfies
\begin{equation}
1-\cos(2\sqrt{l!}t)=\varepsilon_{0},
\label{zeroth-order}
\end{equation}
for a small positive number $0<\varepsilon_{0}\ll 1$. Second, let us consider the case where $n=1$. Because $D_{1}=(l+1)!$ from Eq.~(\ref{definition-D-n-2}), the time variable $t$ that we want to compute satisfies
\begin{equation}
1-\cos(2\sqrt{(l+1)!}t)=\varepsilon_{1},
\label{first-order}
\end{equation}
for a small positive number $0<\varepsilon_{1}\ll 1$. From Eq.~(\ref{zeroth-order}), we obtain
\begin{equation}
t
\simeq
\frac{q}{\sqrt{l!}}\pi
\quad
\mbox{for $q=0,1,2,...$.}
\label{zeroth-order-time}
\end{equation}
Moreover, substituting Eq.~(\ref{zeroth-order-time}) into Eq.~(\ref{first-order}) yields
\begin{equation}
\cos(2\sqrt{l+1}q\pi)
\simeq
1-\varepsilon_{1}.
\end{equation}
Thus, we arrive at
\begin{eqnarray}
|2\sqrt{l+1}q\pi-2p\pi|&<&\delta(\varepsilon_{1})\ll 1,
\label{first-order-a} \\
|2\sqrt{l+1}q\pi-4p\pi|&<&\delta(\varepsilon_{1})\ll 1,
\label{first-order-b} \\
|2\sqrt{l+1}q\pi-6p\pi|&<&\delta(\varepsilon_{1})\ll 1,
\label{first-order-c} \\
...&& \nonumber
\end{eqnarray}
where $\delta(\varepsilon_{1})=|\arccos(1-\varepsilon_{1})|$.

For Eqs.~(\ref{zeroth-order}) and (\ref{first-order}) to hold, only one of Eqs.~(\ref{first-order-a}), (\ref{first-order-b}), (\ref{first-order-c}), ... must be satisfied for a specific positive integer $p$ that is coprime with $q$. Thus, we must find positive integers $q$ and $p$ such that
\begin{equation}
\left|
\frac{\sqrt{l+1}}{k}-\frac{p}{q}
\right|
<
\frac{\delta(\varepsilon_{1})}{2kq\pi},
\end{equation}
for a positive integer $k=1,2,3,...$. Summarizing the above, we want to find positive integers $p$ and $q$ where
\begin{equation}
\left|
\frac{\sqrt{l+1}}{k}-\frac{p}{q}
\right|
=
\frac{1}{cq^{1+\nu}}
\quad
\mbox{for $k=1,2,3,...$},
\label{Diophantine-approximation-1-nu}
\end{equation}
with constant values $c$ and $\nu>1$.

Here, we introduce the following theorem that relates to the Diophantine approximation \cite{Coppel2009,Niven1960}.
For an arbitrary irrational number $\alpha$, there exists infinite sequences $p_{m}$ and $q_{m}(>0)$ for $m\geq 0$ such that $p_{m}$ and $q_{m}$ are coprime and
\begin{equation}
\left|
\alpha-\frac{p_{m}}{q_{m}}
\right|
<
\frac{1}{q_{m}^{2}}.
\end{equation}
In general, it is known that a computation of the first few terms of a continued fraction representation of an irrational number gives its Diophantine approximation \cite{Coppel2009,Niven1960}. Describing the continued fraction of an arbitrary irrational number $\alpha$ as
\begin{eqnarray}
\alpha
&=&
a_{0}
+\frac{1}{\displaystyle a_{1}
+\frac{1}{\displaystyle a_{2}+ \frac{1}{\ddots}}} \nonumber \\
&=&
[a_{0};a_{1},a_{2},...],
\label{continued-fraction-definition-0}
\end{eqnarray}
where $a_{0}$ is an integer and $a_{1}$, $a_{2}$, ... are positive integers, we can define infinite sequences of integers $(p_{0},p_{1},p_{2},...)$ and $(q_{0},q_{1},q_{2},...)$ such that
\begin{equation}
\frac{p_{m}}{q_{m}}
=
[a_{0};a_{1},a_{2},...,a_{m}].
\end{equation}
Then, the following relationships hold:
\begin{equation}
\left|
\alpha-\frac{p_{m}}{q_{m}}
\right|
\leq
\frac{1}{a_{m+1}q_{m}^{2}}
\quad
\mbox{for $m=0,1,2,...$.}
\end{equation}

Using the above, we compute $t=q\pi$ that satisfies $|S_{z}(t)|\ll 1$ for $l=2$, i.e., the two-photon JCM. In particular, we want to find $(q_{m},p_{m})$ for
\begin{equation}
\left|
\frac{\sqrt{3}}{k}-\frac{p_{m}}{q_{m}}
\right|
<
\frac{1}{q_{m}^{2}}.
\end{equation}
Here, we define the following convenient notation that represents the continued fraction of an irrational number $\alpha$:
\begin{equation}
\chi_{\alpha}(m)
=
[a_{0};a_{1},a_{2},...,a_{m}]
\quad
\mbox{for $\alpha=[a_{0};a_{1},a_{2},a_{3},...]$.}
\end{equation}
Furthermore,
\begin{equation}
\sqrt{3}
=
[1;1,2,1,2,...],
\end{equation}
\begin{equation}
\chi_{\sqrt{3}}(12)=\frac{3691}{2131},
\quad
\chi_{\sqrt{3}}(13)=\frac{5042}{2911},
\quad
\chi_{\sqrt{3}}(14)=\frac{13{\,}775}{7953},
\quad
....
\end{equation}
Let us consider the following set of denominators of $\chi_{\sqrt{3}}(12)$, $\chi_{\sqrt{3}}(13)$, ..., $\chi_{\sqrt{3}}(59)$: \\${\cal M}_{\sqrt{3}}=\{2131,2911,7953, ...\}$. Here, we define
\begin{eqnarray}
{\cal M}_{\sqrt{3}/k}
&=&
\{
\mbox{denominators of }
\chi_{\sqrt{3}/k}(12), \chi_{\sqrt{3}/k}(13), ..., \chi_{\sqrt{3}/k}(59)
\} \nonumber \\
&&
\quad
\mbox{for $k=2,3,4,5,6,7$}.
\end{eqnarray}
Next, we take the union of these sets:
\begin{equation}
{\cal M}=
{\cal M}_{\sqrt{3}}
\cup
{\cal M}_{\sqrt{3}/2}
\cup
{\cal M}_{\sqrt{3}/3}
\cup
\cdots
\cup
{\cal M}_{\sqrt{3}/7}
\cup
\{0\}.
\end{equation}
The number of elements of ${\cal M}$ is $243$. By rewriting Eq.~(\ref{L-4-t}) as
\begin{equation}
L^{(3)}(t)
=
-\frac{1}{2}(1-b^{l})[1-(1-b)f(t)],
\end{equation}
\begin{equation}
f(t)
=
\sum_{n=0}^{\infty}b^{n}\cos(2\sqrt{D_{n}}t),
\end{equation}
where $b=e^{-\beta}$ and $-\epsilon<L^{(3)}(t)<\epsilon$, we obtain
\begin{equation}
-
\frac{2\epsilon}{(1-b)(1-b^{l})}
+
\frac{1}{1-b}
<
f(t).
\label{inequality-f-t}
\end{equation}
Moreover, up to $O(b^{2})$ for $l=2$, we obtain
\begin{eqnarray}
&&
(1-2\epsilon)
+
(1-2\epsilon)b
+
(1-4\epsilon)b^{2} \nonumber \\
&<&
\cos(2\sqrt{D_{0}}t)
+
b\cos(2\sqrt{D_{1}}t)
+
b^{2}\cos(2\sqrt{D_{2}}t),
\label{second-order-inequality}
\end{eqnarray}
where $D_{0}=2$, $D_{1}=6$, and $D_{2}=12$. The subset of ${\cal M}$ whose elements satisfy Eq.~(\ref{second-order-inequality}) with $\beta=2.0$ and $\epsilon=0.05$ is given by
\begin{equation}
\tilde{\cal M}
=
\{0,15{\,}731{\,}042,1{\,}117{\,}014{\,}753, ...\}.
\label{tilde-M-l-2}
\end{equation}
The number of elements of $\tilde{\cal M}$ is equal to $15$. The blue curves in Fig.~\ref{figure06}(b) plot $(\beta,S_{x}(t))$ for $t=q\pi$ and $q\in\tilde{\cal M}$. Here, the blue curves pass through red points, but not all of them. Thus, we must admit that $\tilde{\cal M}$ gives us only part of the whole $t$ such that $|S_{z}(t)|<\epsilon$. We carried out similar calculations for cases with $l=1$, $3$, and $4$ and draw their curves in Fig.~\ref{figure06}(a), (c), and (d). These figures tell us that the $\tilde{\cal M}$ obtained by the Diophantine approximation covers only some of the red dots. The calculations are described in Appendix~\ref{section-appendix-B}.

Here, we explain why the graphs of Fig.~\ref{figure03}(a) and (b) are not symmetrical with respect to the $S_{z}$-axis. In Appendix~\ref{section-appendix-B}, we show that we do not need to apply the Diophantine approximation to Eq.~(\ref{Diophantine-approximation-1-nu}) for $l=3$. Up to $O(b)$ for $l=3$, $L^{(3)}(t)=0$ holds for $t=q\pi/\sqrt{6}$, where $q=0,1,2,3,...$. Now, we expand $L^{(1)}(t)$ given by Eqs.~(\ref{A-elements-0}) and (\ref{L1234-definition}) as
\begin{eqnarray}
L^{(1)}(t)
&=&
\cos(\sqrt{D_{0}}t)\cos(\sqrt{D'_{0}}t) \nonumber \\
&&
-
[
\cos(\sqrt{D_{0}}t)\cos(\sqrt{D'_{0}}t)
-
\cos(\sqrt{D_{1}}t)\cos(\sqrt{D'_{1}}t)]b+O(b^{2}).
\label{L1t-expansion-b}
\end{eqnarray}
From Eqs.~(\ref{definition-D-n}) and (\ref{definition-D-dash-n}), we obtain $D_{0}=6$, $D_{1}=24$, and $D'_{0}=D'_{1}=0$ for $l=3$. Thus, substitution of $t=q\pi/\sqrt{6}$ into Eq.~(\ref{L1t-expansion-b}) gives
\begin{eqnarray}
L^{(1)}(q\pi/\sqrt{6})
&=&
(-1)^{q}-[(-1)^{q}-1]b+O(b^{2}) \nonumber \\
&=&
\left\{
\begin{array}{ll}
1+O(b^{2}) & \mbox{for $q=0,2,4,...$}, \\
-1+2b+O(b^{2}) & \mbox{for $q=1,3,5,...$}.
\end{array}
\right.
\end{eqnarray}
Hence, if we consider the low-temperature limit (but not absolute zero, i.e., $b\neq 0$) up to $O(b)$ with $|S_{z}|\ll 1$ for $l=3$, the plot of points $(S_{x},S_{z})$ is not symmetric under the inversion $S_{x}\to -S_{x}$. This fact allows the trajectories of $\mbox{\boldmath $S$}(t)$ to break the invariance. This situation occurs when $l=3$, but not when $l=1,2$ or $4$.

Drawing the trajectories in Fig.~\ref{figure03}(a) and (b), we set $\mbox{\boldmath $S$}(0)=(1,0,0)^{\text T}$ as the initial state of the Bloch vector. Then, $\mbox{\boldmath $S$}(t)$ $\forall t>0$ is given by Eq.~(\ref{Bloch-vector-xz-plane-0}). By contrast, if we set $\mbox{\boldmath $S$}(0)=(-1,0,0)^{\text T}$ as the initial state, we obtain $\mbox{\boldmath $S$}(t)=(-L^{(1)}(t),0,L^{(3)}(t))^{\text T}$ $\forall t>0$ and the sign of $S_{x}(t)$ is reversed. Hence, considering those dual trajectories of $\mbox{\boldmath $S$}(t)$ together, we can recover their invariance under the inversion of $S_{x}\to -S_{x}$.

Finally, we should remark that the trajectories of $\mbox{\boldmath $S$}(t)$ are invariant under the inversion of $S_{x}\to -S_{x}$ for $l=1,2$, and $4$ in Figs.~\ref{figure02}, \ref{figure03}, and \ref{figure05}, although we set $\mbox{\boldmath $S$}(0)=(1,0,0)^{\text T}$ as the initial state. This is because the trajectories are quasiperiodic and there exists a time $t_{\text {large}}(\gg 1)$ in the distant future where $\mbox{\boldmath $S$}(t_{\text {large}})\simeq (-1,0,0)^{\text T}$ holds.

\section{\label{section-conclusion-discussion}Conclusion}
We studied the discrete trajectories of the Bloch vector of multiphoton JCM at finite temperatures. First, we showed that their graphs are invariant under scale transformations of the time variable. Second, we showed that we can compute the times at which the absolute value of the $z$-component of the Bloch vector is nearly equal to zero in the Diophantine approximation. The trajectories of the Bloch vector for a continuous time variable are disordered and show completely random behavior because of thermal effects. In contrast, if we plot the Bloch vector at constant time intervals, such that the time variable takes discrete values at $t_{n}=n\Delta t$ for $n=0, 1, 2, ..., N$, the graphs of the trajectories reveal regularities.

The reason for the above findings is that, in the zero-temperature limit, $\beta\to\infty$, $e^{-n\beta}$ in Eq.~(\ref{L-4-t}) vanishes for $n\neq 0$, whereas at finite low-temperature not equal to zero, $e^{-n\beta}$ cannot be neglected for $n\neq 0$ and $L^{(3)}(t)$ approximates to a finite sum of trigonometric functions whose dimensionless angular frequencies are irrational numbers. A similar situation occurs for $L^{(1)}(t)$. Hence, the quasiperiodicity of the Bloch vector arises because of thermal effects.
The quasiperiodicity causes long-range correlations in time and leads to plots of the Bloch vectors being distributed as shown in Figs.~\ref{figure02}, \ref{figure03}, \ref{figure04}, and \ref{figure05}.

The scale invariance and the Diophantine approximation are unexpected aspects of multiphoton JCMs. We think that the original and multiphoton JCMs still have unknown characteristics that should be explored.

\appendix

\section{\label{section-appendix-A}Implementation of the two-photon JCM with
\\
Josephson junctions}

\begin{figure}
\begin{center}
\includegraphics[width=0.45\linewidth]{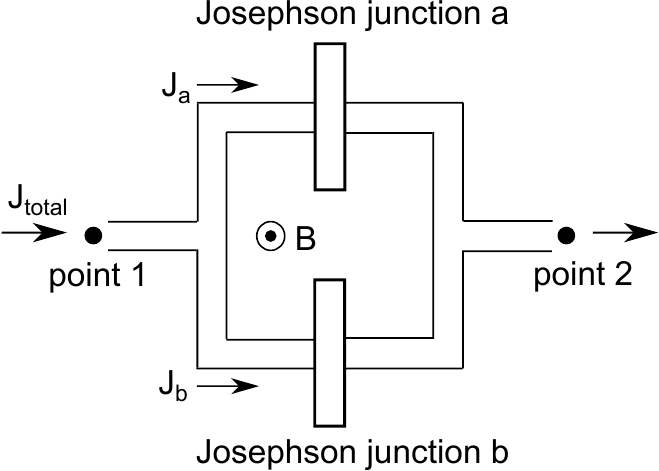}
\end{center}
\caption{A circuit with two Josephson junctions in parallel. The magnetic field pierces the plane of the circuit.}
\label{figure07}
\end{figure}

In this section, we explain how to realize the two-photon JCM experimentally. As shown in Fig.~\ref{figure07}, we consider two Josephson junctions in parallel and set a magnetic field $\mbox{\boldmath $B$}$ that points perpendicularly out of the plane of the circuit in the center \cite{Felicetti2018,Kittel2005}. We denote the phase difference of a wave function between points $1$ and $2$ along the path of Josephson junction $a$ as $\delta_{a}$ and the phase difference between points $1$ and $2$ along the path of Josephson junction $b$ as $\delta_{b}$. Furthermore, we denote the magnetic flux of the magnetic field $\mbox{\boldmath $B$}$ as $\Phi$. Accordingly, we have
\begin{equation}
\delta_{a}
=
\delta_{0}
-
\frac{e}{\hbar c}\Phi,
\quad
\delta_{b}
=
\delta_{0}
+
\frac{e}{\hbar c}\Phi,
\end{equation}
where $\delta_{0}$ is a constant and we have written $\hbar$ explicitly. We denote the total current, current along junction $a$, and current along junction $b$ as $J_{\text{total}}$, $J_{a}$, and $J_{b}$, respectively.

Because $J_{a}$ and $J_{b}$ are currents induced by the Josephson effect, we have
\begin{eqnarray}
J_{\text{total}}
&=&
J_{a}+J_{b} \nonumber \\
&=&
J_{0}(\sin\delta_{a}+\sin\delta_{b}) \nonumber \\
&=&
2J_{0}\sin\delta_{0}\cos\frac{e\Phi}{\hbar c},
\end{eqnarray}
where $J_{0}$ is the maximum zero-voltage current that can be passed by the junctions. Moreover, because
\begin{equation}
J_{0}
\propto
\sigma_{z}
=
|L\rangle\langle L|
-
|R\rangle\langle R|,
\end{equation}
where $|L\rangle$ and $|R\rangle$ are the left- and right-circulating persistent current states, and
\begin{equation}
\Phi
\propto
a+a^{\dagger},
\label{phase-operator}
\end{equation}
we obtain
\begin{eqnarray}
J_{\text{total}}
&\propto&
\sigma_{z}\cos(a+a^{\dagger}) \nonumber \\
&=&
\sigma_{z}+\frac{1}{2}\sigma_{z}(a+a^{\dagger})^{2}+... .
\end{eqnarray}
We can write Eq.~(\ref{phase-operator}) because $\langle\alpha|(a+a^{\dagger})|\alpha\rangle=2\mbox{Re}[\alpha]$, where the magnetic field $\mbox{\boldmath $B$}$ is represented as a coherent state $|\alpha\rangle$ with phase $\alpha$.

Here, we regard $\{|L\rangle,|R\rangle\}$ as a two-level system. Next, by giving a bias to the Josephson junctions, we can transform the basis vectors of the two-level system,
\begin{equation}
|0\rangle
=
\frac{1}{\sqrt{2}}(|L\rangle +|R\rangle),
\quad
|1\rangle
=
\frac{1}{\sqrt{2}}(|L\rangle -|R\rangle),
\end{equation}
and we can replace the operator $\sigma_{z}$ with $\sigma_{x}$. Accordingly, we obtain
\begin{equation}
J_{\text{total}}
\propto
\sigma_{x}+\frac{1}{2}\sigma_{x}(a+a^{\dagger})^{2}+... .
\end{equation}
The second term of the above equation is a Hamiltonian of the two-photon quantum Rabi model. It can be rewritten as
\begin{equation}
\frac{1}{2}\sigma_{x}(a+a^{\dagger})^{2}
=
\frac{1}{2}
(\sigma_{+}+\sigma_{-})(a+a^{\dagger})^{2}
\end{equation}
and apply the rotating-wave approximation to it to attain finally the interaction term of the two-photon JCM $\sigma_{+}a^{2}+\sigma_{-}(a^{\dagger})^{2}$.

\section{\label{section-appendix-B}Calculations for drawing the curves in
\\
Fig.~\ref{figure06}(a), (c), and (d)}
In this section, we explain how to calculate the curves drawn in Fig.~\ref{figure06}(a), (c), and (d). For $l=1$, that is, the original JCM, we compute the following continued fractions:
\begin{equation}
\sqrt{2}
=
[1;2,2,2,2, ...],
\end{equation}
\begin{equation}
\chi_{\sqrt{2}}(12)
=
\frac{47{\,}321}{33{\,}461},
\quad
\chi_{\sqrt{2}}(13)
=
\frac{114{\,}243}{80{\,}782},
\quad
\chi_{\sqrt{2}}(14)
=
\frac{275{\,}807}{195{\,}025},
\quad
....
\end{equation}
We consider a subset of denominators of $\chi_{\sqrt{2}}(12)$, $\chi_{\sqrt{2}}(13)$, ..., $\chi_{\sqrt{2}}(39)$,
\begin{equation}
{\cal M}_{\sqrt{2}}
=
\{
33{\,}461, 80{\,}782, 195{\,}025, ...\}.
\end{equation}
Similarly, we define
\begin{eqnarray}
{\cal M}_{\sqrt{2}/k}
&=&
\{
\mbox{denominators of }
\chi_{\sqrt{2}/k}(12), \chi_{\sqrt{2}/k}(13), ..., \chi_{\sqrt{2}/k}(39)
\} \nonumber \\
&&
\quad
\mbox{for $k=2,3,4$}.
\end{eqnarray}
Then, we take a union of the above sets,
\begin{equation}
{\cal M}=
{\cal M}_{\sqrt{2}}
\cup
{\cal M}_{\sqrt{2}/2}
\cup
{\cal M}_{\sqrt{2}/3}
\cup
{\cal M}_{\sqrt{2}/4}
\cup
\{0\},
\end{equation}
where the number of elements of ${\cal M}$ is $91$. Expanding Eq.~(\ref{inequality-f-t}) up to $O(b^{2})$ for $l=1$ gives
\begin{equation}
(1-2\epsilon)
+
(1-4\epsilon)b
+
(1-6\epsilon)b^{2}
<
\cos(2t)
+
b\cos(2\sqrt{2}t)
+
b^{2}\cos(2\sqrt{3}t).
\label{second-order-inequality-l1}
\end{equation}
Taking $\beta=2.0$ and $\epsilon=0.0035$, we obtain the subset of ${\cal M}$ whose elements satisfy Eq.~(\ref{second-order-inequality-l1}) in the form,
\begin{equation}
\tilde{\cal M}
=
\{0,19{\,}601,33{\,}461,470{\,}832,...\}.
\end{equation}
The number of elements of $\tilde{\cal M}$ is equal to $24$. Fig.~\ref{figure06}(a) plots $(\beta,S_{x}(t))$ for $t=q\pi$ and $q\in\tilde{\cal M}$ (blue curves).

For $l=4$, that is, for the four-photon JCM, we compute the following continued fractions:
\begin{eqnarray}
{\cal M}_{\sqrt{5}/k}
&=&
\{
\mbox{denominators of }
\chi_{\sqrt{5}/k}(8), \chi_{\sqrt{5}/k}(9), ..., \chi_{\sqrt{5}/k}(59)
\} \nonumber \\
&&
\quad
\mbox{for $k=1,2,...,8$}, \nonumber \\
{\cal M}_{\sqrt{5}/9}
&=&
\{
\mbox{denominators of }
\chi_{\sqrt{5}/9}(8), \chi_{\sqrt{5}/9}(9), ..., \chi_{\sqrt{5}/9}(58)\},
\end{eqnarray}
\begin{equation}
{\cal M}=
{\cal M}_{\sqrt{5}}
\cup
{\cal M}_{\sqrt{5}/2}
\cup
\cdots
\cup
{\cal M}_{\sqrt{5}/9}
\cup
\{0\},
\end{equation}
where the number of elements of ${\cal M}$ is $323$. [The denominator of $\chi_{\sqrt{5}/9}(59)$ is so huge that computing the continued fraction is not tractable.] Expanding Eq.~(\ref{inequality-f-t}) up to $O(b^{2})$ for $l=4$ gives
\begin{eqnarray}
&&
(1-2\epsilon)
+
(1-2\epsilon)b
+
(1-2\epsilon)b^{2} \nonumber \\
&<&
\cos(2\sqrt{24}t)
+
b\cos(2\sqrt{120}t)
+
b^{2}\cos(2\sqrt{360}t).
\label{second-order-inequality-l4}
\end{eqnarray}
Taking $\beta=2.0$ and $\epsilon=0.04$, we obtain the subset of ${\cal M}$ whose elements satisfy Eq.~(\ref{second-order-inequality-l4}) as $\tilde{\cal M}$ and whose number is equal to $17$. The blue curves in Fig.~\ref{figure06}(d) plot $\tilde{\cal M}$.

For $l=3$, that is, the three-photon JCM, the situation is different from the above. Because $\sqrt{l+1}/k=2/k$ for Eq.~(\ref{Diophantine-approximation-1-nu}), we do not need to compute the Diophantine approximation. Thus, we prepare ${\cal M}=\{0,1,2,...,2000\}$ by letting $q\in{\cal M}$ satisfy one of Eqs.~(\ref{first-order-a}), (\ref{first-order-b}), (\ref{first-order-c}), .... Expanding Eq.~(\ref{inequality-f-t}) up to $O(b^{2})$ for $l=3$ yields
\begin{eqnarray}
&&
(1-2\epsilon)
+
(1-2\epsilon)b
+
(1-2\epsilon)b^{2} \nonumber \\
&<&
\cos(2\sqrt{6}t)
+
b\cos(2\sqrt{24}t)
+
b^{2}\cos(2\sqrt{60}t).
\label{second-order-inequality-l3}
\end{eqnarray}
Taking $\beta=2.0$ and $\epsilon=0.003$, we obtain the subset of ${\cal M}$ whose elements satisfy Eq.~(\ref{second-order-inequality-l3}) as $\tilde{\cal M}$ and whose number is equal to $15$. The blue curves in Fig.~\ref{figure06}(c) plot $\tilde{\cal M}$.

\section*{Acknowledgment}
This work was supported by the MEXT Quantum Leap Flagship Program, Grant No. JPMXS0120351339.

\end{document}